\documentclass{emulateapj}
\shortauthors{Zaritsky \& Harris}
\shorttitle{Drivers of Star Formation}

\begin{document}
\title{Quantifying the Drivers of Star Formation on Galactic Scales. I. The Small Magellanic Cloud}
  
\author{Dennis Zaritsky}
\affil{Steward Observatory, University of Arizona, 933
   N. Cherry Ave., Tucson, AZ 85721, USA; dzaritsky@as.arizona.edu}
\author{Jason Harris\footnote{Current address: Steward Observatory, University of Arizona, 933 N Cherry Ave., Tucson, AZ, 85721, USA; jharris@as.arizona.edu}}
\affil{Space Telescope Science Institute, 3700 San Martin Dr., Baltimore,
MD, 21218, USA}

\begin{abstract} 
We use the star formation history of the Small Magellanic Cloud (SMC) to 
place quantitative limits on the effect of tidal interactions and 
gas infall on the star formation and chemical enrichment history of the SMC.
The coincident timing of two recent  ($<$ 4 Gyr) increases in the star formation rate 
and SMC/Milky Way(MW) pericenter passages 
suggests that global star formation in the SMC is driven at least in part
by tidal forces due to the MW. The Large Magellanic Cloud (LMC) is the other
potential driver of star formation, but is only near the SMC during the most
recent burst. The poorly constrained LMC-SMC orbit is our principal uncertainty.
To explore the correspondence between bursts and MW pericenter passages further,
we model star formation in the SMC using a combination of
continuous and  tidally-triggered star formation.
The behavior of the tidally-triggered mode
is a strong inverse function of the SMC-MW separation (preferred behavior
$\sim r^{-5}$, resulting in a factor of $\sim$ 100 difference in the rate of tidally-triggered
star formation at pericenter and apocenter). Despite the success of these closed-box 
evolutionary models in reproducing the recent SMC star formation history and current 
chemical abundance, 
they have some systematic shortcomings that are remedied by postulating that
a sizable infall event ($\sim$ 50\% of the total gas mass)  occurred $\sim$ 4 Gyr
ago. Regardless of whether this infall event is included, the fraction of stars in the SMC
that formed via a tidally triggered mode is $>$ 10\% and could be as large as $70\%$.

\end{abstract}

\keywords{galaxies: evolution --- galaxies: interactions --- Magellanic Clouds}

\section{Introduction}

Galaxy evolution is sufficiently complex and the processes
that drive star formation sufficiently unknown that current
state-of-the-art models must resort to simple parameterizations of the 
physics involved (see \cite{kauffmann, cole, somerville}). 
Fits of these models to global averages
such as the galaxy luminosity function or volume averaged star formation as a function
of lookback time (the ``Madau" plot; \cite{madau}) provide weak
constraints on important details buried well within the models. 
With the nearly unlimited freedom in 
the parameterization of galaxy interactions and the effects of these interactions on the 
star formation history of galaxies, the experiments as currently
practiced primarily provide consistency checks that are reassuring
but not definitive.  
Progress in this field requires external, empirical, quantitative
constraints on how star formation is driven by
interactions and infall.

Local Group (LG) galaxies are the only galaxies for which current 
observations can resolve sufficiently faint stars, red giant branch or fainter, with which one
can constrain  both the recent and ancient 
star formation history (SFH).  The availability 
of the entire SFH enables us to identify and quantify the factors that
have influenced the star formation rate (SFR) over time.
Even among LG galaxies, such work remains quite difficult
beyond the sphere of influence of the Milky Way (MW)  because color-magnitude
diagrams (CMDs) do not reach stars below the horizontal branch across the entire
face of a galaxy.
Within the accessible volume
there are a number of interesting stellar systems and 
various studies have already illustrated the rich taxonomy 
of SFHs (for a review see \cite{mateo}). The two most massive systems 
(excluding the MW for which study is complicated by our position within it) are the 
Large and Small Magellanic Clouds (LMC and SMC). These satellite galaxies 
are qualitatively different than the remainder of the sample because they currently
contain large gas reservoirs and continue to vigorously form stars. 
Of the systems we can study in detail and in their entirety, these are the
most similar to galaxies observed outside the Local Group.

In addition to its fortuitous placement, 
the Small Magellanic Cloud has another characteristic that
makes it suitable for this study. 
Because it has no spiral structure and is unbarred \citep{zar00},
it has no apparent internal mechanism that can drive global modes of star formation.
More massive galaxies, like the LMC which  has both bars and spiral arms,
are more complex.
Unfortunately, the SMC's low mass makes it susceptible to losing substantial
amounts of gas during its evolution.
Its velocity dispersion is sufficiently low (25.3 km sec$^{-1}$
measured from planetary nebulae  \citep{dopita} and 27 km sec$^{-1}$ from C stars 
\citep{hardy}) that gas outflow during vigorous phases of star
formation is expected \citep{martin, garnett}.  Although we want to understand global
star formation in both
more and less massive systems, the more massive ones will be complicated, 
perhaps dominated, 
by internal drivers of star formation such as local instabilities \citep{spitzer, quirk,
k98} and the less massive ones will be even more susceptible to outflow. The
SMC may be a good initial object for study.

The star formation history of the SMC \citep{harris03}
has several distinctive features that should provide leverage 
on constraining simple evolutionary models. We address the 
following questions: 1) by what factor is star formation increased
during tidal interactions, 2) what fraction of the stars formed
during those tidal interactions, 3) does a closed-box model
satisfactorily reproduce the data, and 4) if infall or outflow is required, 
what are its effects and how much is necessary? In 
\S2 we provide a brief description of the data and refer
readers to \cite{zar02} and \cite{harris03} for more details. In \S3
we describe the models and discuss our fitting of these models
to the SFH of the SMC.
We summarize
and conclude in \S4.

\section{The Data}

The original data come from the Magellanic Clouds Photometric Survey and
are presented and made public by \cite{zar02}. They consist of UBVI photometry of 
over 5 million stars in the central $4^\circ \times 4.5^\circ$ area
of the Small Magellanic Cloud. 
The reconstruction of the star formation history
was done by \cite{harris03} using the StarFISH algorithm 
presented and made public by \cite{harris01}.

\begin{figure}
\begin{center}
\plottwo{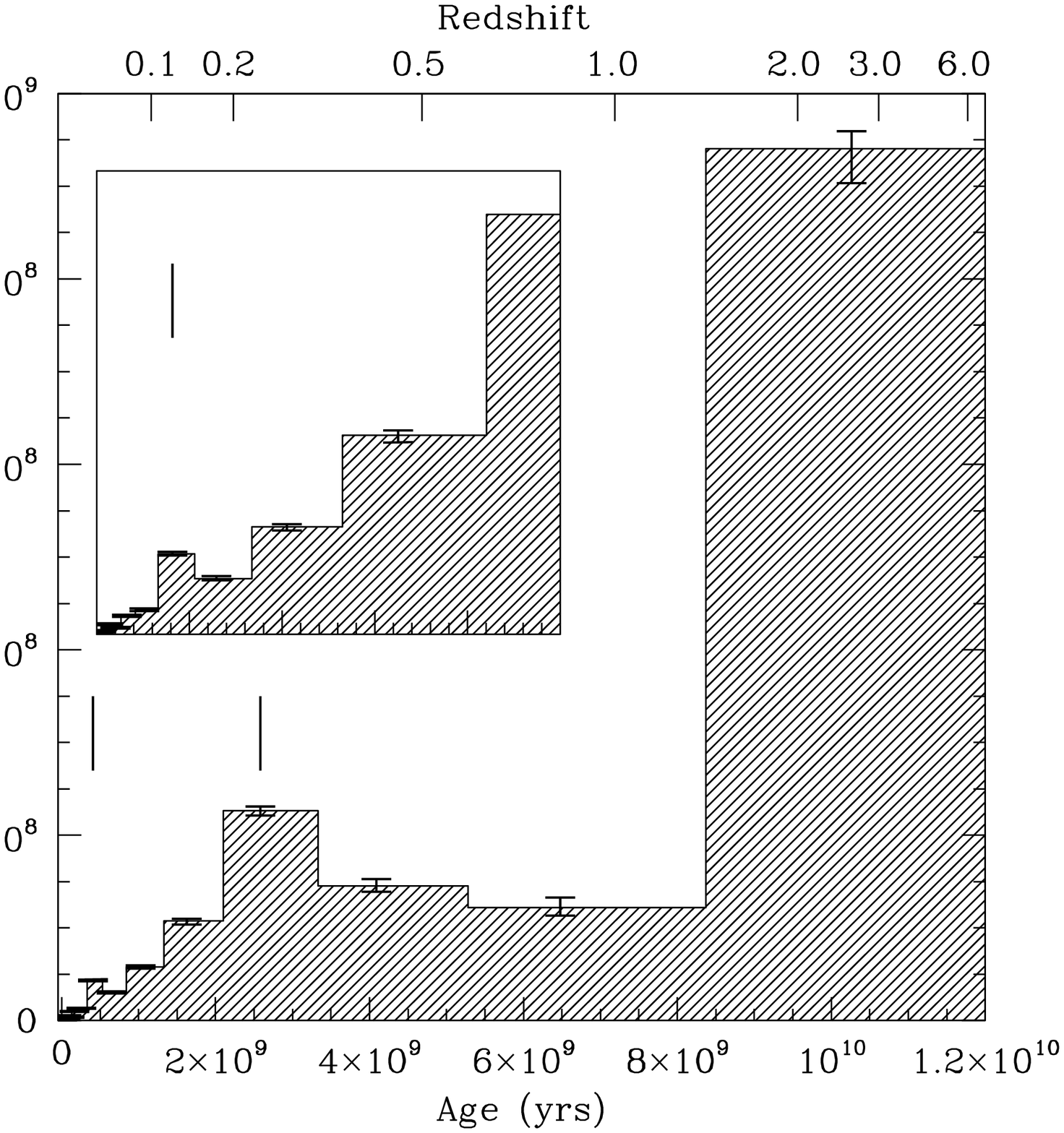}{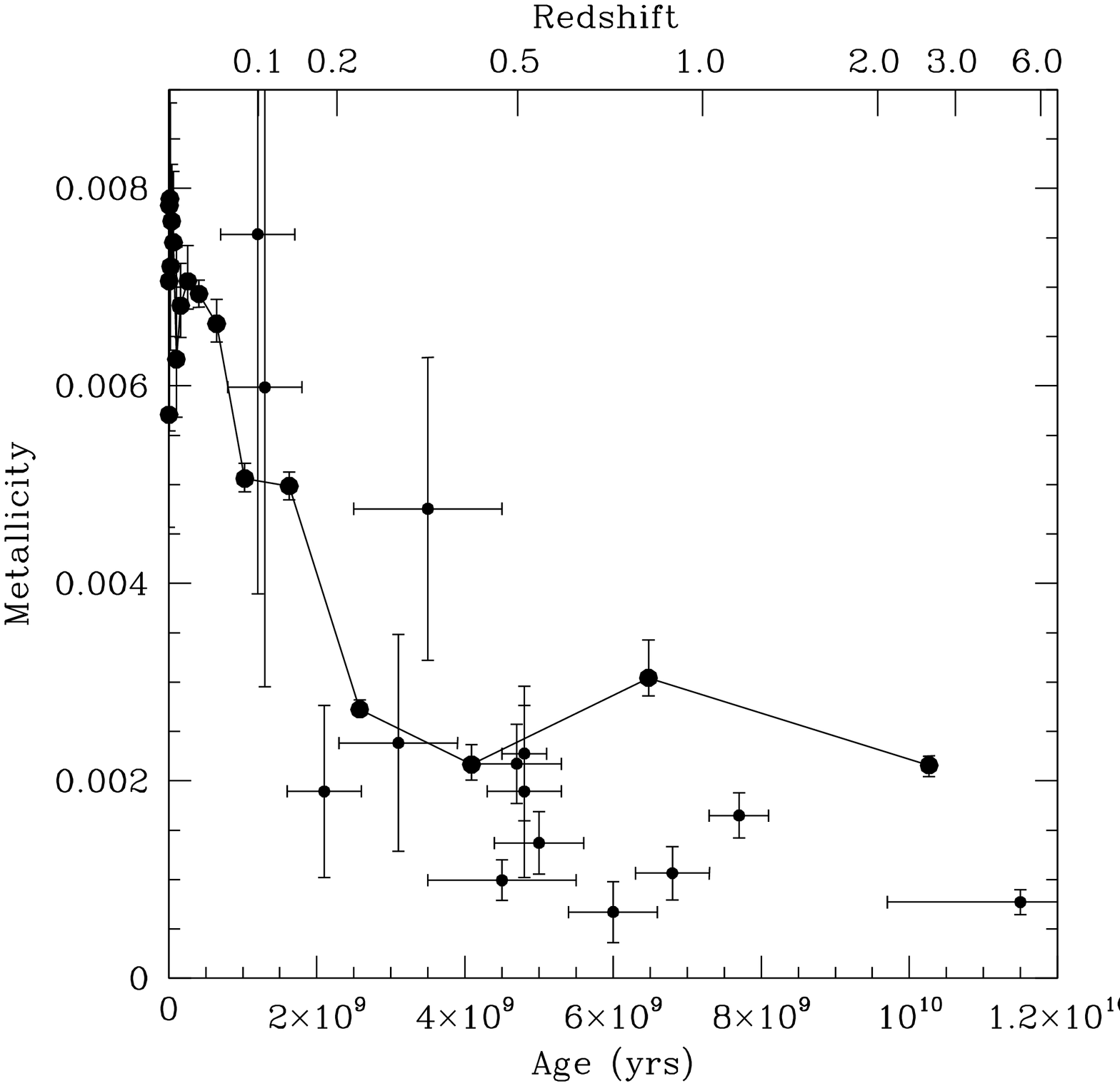}
\caption{Reprise of the results from \cite{harris03}. The left panel 
shows the stellar age function (SAF) for the entire SMC, while the right panel 
shows the chemical enrichment history. Age $= 0$ corresponds to the current
day. The histogram is used only to highlight the shape of the SAF, the height
rather than the integral of each bin represents the number of stars of that
age.
Unconnected circles in the right panel
represent the ages and metallicities of individual clusters \cite{def,piatti}. 
The connected circles represent the results from the field star analysis of 
\cite{harris03}. The inset
in the left panel is a magnification of the plot at young ages. The conversion
from age to redshift in this Figure and throughout the paper is done for 
the WMAP cosmology \citep{spergs}. 
\label{oldresults}}
\end{center}
\end{figure}

We reprise the recovered star formation and chemical enrichment
history in Figure \ref{oldresults}. 
Instead of plotting the star formation rate as a function of time,
which when plotted in age bins of different widths
can be visually misleading if the SFH is bursty,  we plot
the number of stars of each age. 
We refer to this quantity as the stellar age function (SAF). The data
necessary to produce the more standard average SFR plot is
presented by \cite{harris03}.

We  combine the data from all of the spatial bins described by
\cite{harris03} and propagate the internal errors quoted there to 
arrive at the global SAF. Before discussing the SAF, we note that
the oldest bin represents all stars older than $\sim$ 8 Gyr. Therefore, 
although the SAF is well defined for ages $>$ 8 Gyr, the SFR is not.
From  the left panel of 
Figure \ref{oldresults} we draw two key conclusions regarding the SAF:
(1) there are two statistically significant upward 
deviations from the general declining nature of the function with time at
$\sim 0.4$ and 2.5 Gyr and (2) there is clear drop 
in the SAF from the oldest bin to the next oldest (for a constant star formation rate,
the SAF in the $\sim 5$ to 8 Gyr bin would be greater than that
in the $\sim 3.5$ to 5 Gyr bin by 50\%).
The apparent deficit in the 
inferred SFR at intermediate
ages cannot be explained away by extending the width of the bin representing
the oldest stars back in time (see discussion below).

The right panel of 
Figure \ref{oldresults} also contains two intriguing, but less statistically
significant, results regarding
the chemical enrichment: (1) we derive a chemical enrichment
history for field stars from an analysis of broad-band stellar photometry that 
agrees with that from 
the analyses of individual stellar clusters \citep{def, piatti}, and (2) 
the chemical abundance is either constant until $\sim$ 3 Gyr ago or
has a slight decline after $\sim$ 6 Gyr ago.
The latter result is at a fairly low confidence level
from our data alone due to the use of broad-band colors, 
but consistent with the cluster data. Because we consider of the uncertainties we will
not fit our models to the chemical enrichment history, but we will compare
the chemical enrichment history predicted by the models to the observations.

\section{The Models}

We begin our attempts to reproduce the observations
with a simple, closed-box model galaxy evolution model.
The galaxy initially consists of  a certain mass of gas. It converts that gas
to stars at a rate proportional to the current gas mass until the gas is exhausted.
This simple parameterization is 
analogous to the disk star formation law proposed by \cite{schmidt},
SFR $\propto \Sigma^n$ where $\Sigma$ is the gas surface density and $n$ is
to be determined empirically. \cite{schmidt} found
$n = 2$, but more recent work finds $n = 1.3 \pm 0.3$ when the
gas density lies above a threshold value \citep{k89}. Whether these results are directly
applicable to a low-mass system like the SMC, which is not a simple tilted HI disk
\citep{ss}. For simplicity,
and out of ignorance,
we have adopted a SFR $\propto M_g$, which results in an
exponentially declining star formation history.

For the SMC,
where we have both a measurement of the stellar mass, from our
catalog, and the remaining gas mass, from the H I and cold dust measurements of
\cite{stan}, the initial gas mass is observationally constrained to be the sum
of these two masses for a closed-box model. We define
the ``quiescent"  (Q) rate of star formation as the fraction
of the available gas per Gyr that is converted to stars. 
The ``tidally triggered" (TT) mode of 
star formation is modeled with a continually varying SFR that is inversely related 
to the distance between the SMC and MW and is specifically
$A/(r/40 {\rm kpc})^b$, where $r$
is the distance between the SMC and MW in kpc, $A$ is a normalization
constant that sets the relative efficiencies of the Q and TT
modes, and $b$ parameterizes the unknown physics that relates the
tidal field to star formation.
The model assumes complete instantaneous 
recycling of material with effective yield, $y$, which allows us to also 
predict the corresponding age-metallicity relation. We set the yield,
0.009, to reproduce the current chemical abundances. This value is
within the range of effective yields determined for similar galaxies
in previous studies \citep{vila,garnett}. We trace 
the orbit of the SMC for $t_{AGE}$ Gyr and calculate the SFR
along the orbit.
To compare to the observational data,
we calculate  the SAF and stellar metallicity integrated 
over the same time bins defined by isochrone set used by \cite{harris03}.

The available external constraints, such as the measurement of the 
remaining gas mass \citep{stan}, the time of the pericenter passages, \citep{lin, marel}, 
and the current metallicity \citep{vila,garnett},  effectively limit many of the model parameters.
The model has only two parameters, $A$ and $b$, which are unconstrained by 
observations other than the SAF and age-metallicity relationship and which
are somewhat degenerate because they both help determine the relative importance
of the TT mode. 

\subsection{Selecting the Model Parameters}

\subsubsection{Remaining and Initial Gas Masses}

In our closed-box model, 
the galaxy is initially composed entirely of gas, with mass equal to the sum of the present
gaseous and stellar masses. 
From our photometry we have a
measurement of the total number of stars, which is the sum of 
the SAF. For the amount of remaining, unused gas (neutral hydrogen plus molecular gas)
we adopt the measurement provided by
\cite{stan} of $1.2\times10^9 M_\odot$.  The molecular gas mass is highly uncertain, 
but in this current estimate it is about twice that of the H I. 
Our simple model does not differentiate between atomic and molecular gas.
Constraining the initial gas mass fixes the integrated star formation efficiency
because the model must produce the
observed number of stars. 
Once we select the relative
efficiencies of the Q and TT modes, the
model iterates to identify the specific values of those efficiencies that produce
a match to both the final total stellar mass and the remaining gas mass. 
For a typical model of the SMC the inferred
efficiency of the Q mode is such that $\sim 1$\% of the remaining 
gas mass is converted to stars per Gyr.

\subsubsection{Age}

As we discussed previously, we have no empirical constraint on the extent of
the oldest bin backward in time from our data alone. 
The choice of when the SMC begins to form stars sets the ratio
of the number of stars in the oldest bin to the number in 
all the other bins. Selecting an older age
for the SMC's initial star formation
places relatively more stars in the oldest bin, exhausts more of the gas at early times,
and decreases the number of stars in the younger bins. As we run various models
for the SMC
(see below), we find that we are driven to adopt older SMC ages in an attempt to explain the
dramatic difference in the SAF between the oldest and second oldest bins. 
The age that we find provides a good
fit for the SMC (17 Gyr, see below), exceeds the measurement of the age of
the Universe from the WMAP satellite (13.7 $\pm$ 0.2 Gyr, \cite{spergs}). Because the
age of the SMC cannot exceed 13.7 Gyr, this discrepancy reflects the difficulty
that the model has in producing a sufficient number of stars in the oldest bin. 
Given the uncertainties in determining the SAF
at the oldest ages and the realization that the closed-box model is probably least
applicable at early times when the SMC is forming, we do not stress or require fine 
agreement at these times, nor do we interpret the best-fit age literally. Alternatively
we could artificially increase the SFR at early times and adopt the WMAP age.
Other than in this respect, the selection of the initial age does not affect the models.

\begin{figure}
\plotone{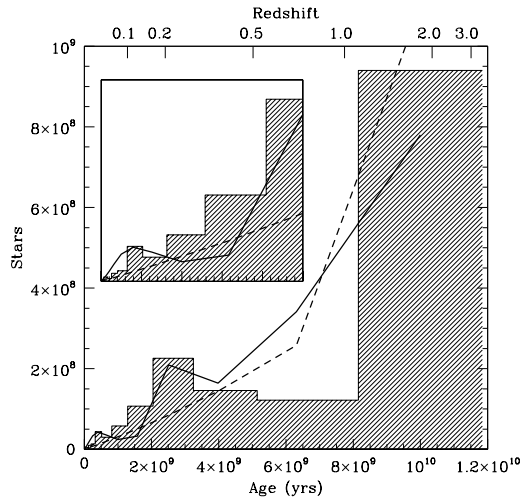}
\caption{Comparison of a model with only quiescent-mode star formation
(dashed line; Model 1 in Table 1), a model with tidally-triggered bursts
(solid line; Model 2 in Table 1), and the measured SAF (histogram).
The inset is an expanded version of the most recent 2.5 Gyr.
\label{bursts}}
\end{figure}

\subsubsection{Orbital Model}

For a model invoking a TT mode of star formation, 
the adopted orbit is critical  because it determines the rate and relative severity of the
pericenter passages. We model the SMC's orbit about the MW using a simple analytic function that
captures the oscillatory nature of the radial behavior
with a {\sl sine} function and the slow decay due to dynamical friction with
a linear decline with time. We fix
the slope of the amplitude decline to obtain a qualitatively
acceptable fit to the data presented in Figures 11 and 12 of \cite{lin}.
The advantage
of this simple description over a more realistic orbit derived by integrating the
equations of motion through
a Galactic potential is that the result of period and phase changes on the SAF can be easily
explored. For a few cases, we compare the resulting SAF using the analytic approach to that from 
full orbit integrations and find no significant difference.

The success of the TT mode in reproducing features in the SAF is first demonstrated in
Figure \ref{bursts}. We compare two models (Models 1 \& 2), the first
of which has zero efficiency in the TT mode, the second of
which has a TT amplitude interactively set to produce a good qualitative fit to the
observed SF peaks at $~\sim 0.4$ and 2.5 Gyr.  Model parameters and reduced $\chi^2$
values are given in Table 1. Model 2 demonstrates that a simple implementation of 
pericenter-driven star formation can reproduce the fluctuations
observed in the SMC's recent ($t < 4$ Gyr) SAF. Model 1, with no TT mode, fails to 
reproduce the observed SAFs at the times of the two bursts by 27 and 35$\sigma$, 
while Model 2 fails by only 3.5 and 1.1$\sigma$. Although the TT mode is not
a unique way to enhance the star formation at these two epochs, the improvement in 
reproducing the SAF is dramatic with even minimal tuning.
We discuss the choice and sensitivity of the results
to the orbital parameters next.

\begin{figure}
\plotone{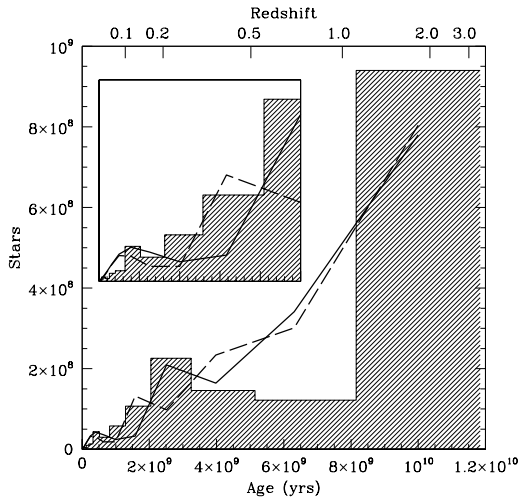}
\caption{Comparison of the same tidally-triggered model as
in Figure \ref{bursts} (solid line; Model 2 in Table 1), a tidally-triggered
burst model with an orbital period of 1.5 Gyr (dashed line; Model 3 in
Table 1), and the measured SAF (histogram).
The inset is an expanded version of the most recent 2.5 Gyr.
\label{period}}
\end{figure}

In this admittedly oversimplified model, the period and phase of the orbit directly determine
the times at which bursts are observed. If there is no discernible lag
between pericenter passage and increase in star formation, then the
peaks in the SAF determine the orbital period and phase. Because we do not know
whether tidally triggered star formation is expected to peak exactly at 
pericenter, the positions of the peaks do not yet place a strong
constraint on the orbital phase.  If we allow the period and phase of the orbit
to vary so as to optimize the match, we find a best-fit period that is slightly longer than that
given by \cite{lin} or \cite{marel} of 2.4 Gyr and a time since the most recent
pericenter of 0.2 to 0.4 Gyr (depending slightly on other parameter
choices). A comparison to a model with an orbital period of 1.5 Gyr (Model 3)
is given in Figure \ref{period} to demonstrate the sensitivity of the predicted
SAF to the chosen period. The 1.5 Gyr period model fails because it does not
reproduce the separation between the two peaks in the SAF.

Is our estimate of the orbital period in conflict with the 
constraints from proper motion studies? The most detailed recent
study of the LMC proper motion comes from the study of \cite{marel}.
If, as is generally seen in simulations \citep{gardiner94}, the SMC and
LMC orbit the MW as a pair, then the parameters of the LMC orbit should
be quite similar to those of the SMC. Determining the orbit of 
the LMC around the MW requires both a precise measurement of its present-day
space velocity and an accurate model of the Galactic potential. We 
model the MW potential with a spherical isothermal dark halo of circular
velocity 180 km sec$^{-1}$ and an interior disk + spheroid mass of 
$1.5 \times 10^{11} M_\odot$. This model has an enclosed mass of $5\times 10^{11}
M_\odot$ at the current position of the LMC (50 kpc), which agrees
with determinations of the enclosed mass at that position \citep{lin,kochanek},
and an enclosed mass of $1.8 \times 10^{12} M_\odot$
at 230 kpc, which agrees
with estimates of the enclosed mass at that radius \citep{zar89,wilk}.
In the most recent determination of the LMC proper motion,
\cite{marel} find that $v_{LMC,rad} = 84\pm 7$ km s$^{-1}$
and $v_{LMC,tan} = 281 \pm 41$ km s$^{-1}$. Using these values
as the initial conditions of an orbit that is integrated backward in the
adopted potential results
in a time since pericenter passage of 0.09 Gyr and a period of 1.9 Gyr.
However, the tangential velocity has a large uncertainty and increasing
it by 1$\sigma$ results in a time since pericenter passage of 0.06 Gyr
and a period of 2.7 Gyr. Therefore, our estimated period of 2.4 Gyr is within
a 1$\sigma$ change in the tangential velocity.

\begin{figure}
\plotone{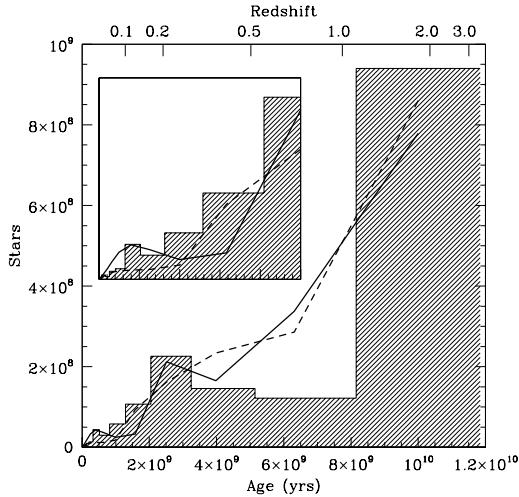}
\caption{Comparison of the same tidally-triggered model as
in Figure \ref{bursts} (solid line; Model 2 in Table 1) and a model
that is identical except the time since pericenter has been changed
to 2.1 Gyr (dashed line; Model 4 in Table 1).
The inset is an expanded version of the most recent 2.5 Gyr.
\label{phasechange}}
\end{figure}

The time since the most recent pericenter passage is less well 
constrained from the star formation modeling because we do not know
whether the star formation maxima correspond
precisely to pericenter passages. One can imagine models in which they
either lag or lead the pericenter passage.
Eventually, when the proper motions and Galactic potential are 
better determined and the orbital constraints are sufficiently tight, this type of
analysis could determine at what phase in the orbit the star formation is
maximized during an interaction. 
For now, we conclude that (1) the peaks are well reproduced given
the published orbital parameters, and (2) the lack of peaks at ages $>$ 4
Gyr is not a result of a lack of a rise in star formation at pericenter, but rather an
artifact of the large binning made necessary by our poorer resolution 
for those times.  The binning, not only the width but also the placement
of the bins, is critical as is shown in
Figure \ref{phasechange} where the model (Model 4), with an adopted time since last pericenter
of 2.1 Gyr,  lacks strong peaks even though it is otherwise 
exactly the same as Model 2.
 This comparison provides a cautionary note that one should not necessarily
interpret smooth star formation histories as evidence against episodic star 
formation (see also \cite{harris03a}) and that one should strive for the highest
temporal resolution possible when investigating triggered bursts. Our bins are determined
by the isochrone set and degeneracies between certain ages given the quality of our
data (see \cite{harris03} used to recover the SFH. The binning was chosen independent
of any observed variations in the SAF. Given a precise model for the star formation history, 
one could create a CMD and compare directly to the data. This method would provide a
more sensitive way to test details of the model, but a more difficult way to explore
parameter space.

\subsubsection{Tidal Triggering}

The crude, basic idea behind tidal triggering is that a rapid change
in the tidal forces that a gas-rich galaxy experiences
should disturb the kinematics of molecular clouds,
hence increase the rate of cloud-cloud collisions, and thereby
increase the star formation rate.  Such models
have been invoked over the previous several decades to qualitatively explain the range
of phenomenon from 
relatively minor star formation enhancements \citep{larson} to
ultraluminous starburst galaxies (for review see \cite{sanders}). The details of how the
increase in star formation scales with tidal force is unknown, but it presumably
scales as some inverse power of the distance between the two galaxies.
To model the dependence of the star formation rate on the separation between
galaxies, $r$, we adopt the relationship that the SFR $\propto 1/r^b$.
We have no external
constraints on the value of $b$ other than it must be
$>$ 0 for star formation to increase with decreasing galaxy separation.
As $b$ increases the tidal effect becomes more
impulsive and the form of the star formation enhancement approaches $\delta(t- t_{peri})$.

\begin{figure}
\plotone{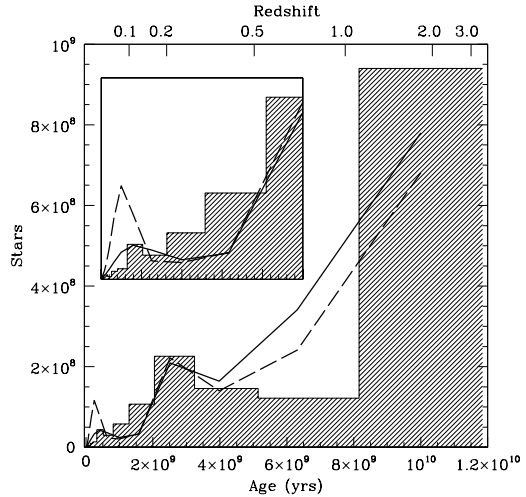}
\caption{Comparison of a tidally-triggered model with $b =  2.4$
(solid line; Model 2 in Table 1), a tidally-triggered model with $b =  10$
(dashed line; Model 5 in Table 1),  and the measured SAF
(histogram). The inset is an expanded version of the most recent 2.5 Gyr.
\label{changeb}}
\end{figure}

\begin{figure}
\plotone{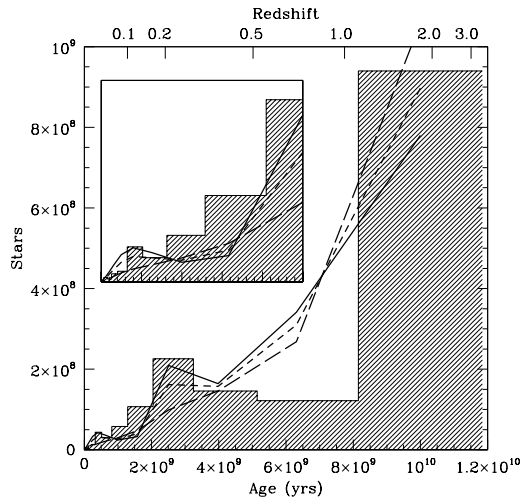}
\caption{Comparison of a tidally-triggered model with $A =  1$
(long-dashed line; Model 7 in Table 1), 
a tidally-triggered model with $A =  10$
(short-dashed line; Model 6 in Table 1),  
a tidally-triggered model with $A =  100$
(solid line; Model 2 in Table 1),  
and the measured SAF
(histogram). The inset is an expanded version of the most recent 2.5 Gyr.
\label{changea}}
\end{figure}

In Model 5  we explore the result of changing $b$ from 2.4 to 10 (Figure \ref{changeb}). 
Even with the much more pronounced bursts, the bursts that occurred
more than 4 Gyr ago  are not discernible with the poorer temporal resolution available
at those times.
A change in $b$ can be countered or enhanced with a
change in the amplitude $A$ of the TT mode (Figure \ref{changea}, Models 6 $-$ 7).
Large changes in $A$ produce modest changes in the SAF, but dominate the
calculation of the fraction of stars formed in the burst mode (for example, changing
A from 10 to 1 reduces the fraction of stars formed in the TT mode from 58\% to 12\%). 

\subsubsection{The Influence of the LMC?}

One obvious ingredient that is missing from this model is the tidal triggering
of star formation by the interaction
between the LMC and SMC. It is much more difficult to model the orbit of the
Clouds about each other, partly because the velocities are smaller so the observational
uncertainties affect the solution to a greater degree and partly because the mass
profiles of the two galaxies are not well known. It is therefore correspondingly
more difficult to connect the SAF to the LMC-SMC orbital parameters. The SMC orbital
models of \cite{gardiner96, yn} find that the most recent LMC-SMC pericenter passage
is nearly coincident with the SMC-MW pericenter passage, 
thereby complicating the identification
of the cause of that burst of star formation. According to the models, the tidal force
exerted by the LMC was significantly larger at the most recent LMC-SMC pericenter
than at any other time during the last 2 Gyr. Because the orbit of the SMC about the LMC is
rather irregular (see Figure 3 of \cite{gardiner96}) it is difficult to predict the times
of any but the most recent close passage, but the models do not 
predict a close LMC pericenter passage
at about 2 Gyr, where we find our older peak. Given the conclusions of \cite{yn} regarding the
effect of the most recent LMC-SMC pericenter passage, one could conclude that a composite model is required 
where the LMC is responsible for the recent burst and the MW for the older burst. With our
limited observational constraints we continue using our simple MW-only model, but 
future work must address the relative importance of the two interactions over the 
age of the SMC rather than just during the most recent passage.

There are several potential tests that could
determine whether the MW or LMC close passages dominate the
SFR. First, if the orbit of the LMC/SMC system about the MW was determined
precisely to have a period of 1.5 Gyr, then we could exclude the MW as the 
dominant influence because of the SAF peak at $\sim$ 2.5 Gyr. As we discussed
above, the current observational constraints are not sufficiently precise to exclude
a period of 2.4 Gyr (necessary to reproduce the SAF). Second, 
if a comparable SAF for the LMC showed the same two bursts as seen in 
the SMC's SAF, then we could rule out the LMC as dominant influence because
the SMC's tidal force should have a much smaller on the LMC than that of the LMC
on the SMC. The available LMC data are currently inconclusive.
Finally, if similar bursts are observed (or can be used to model the SAF) of other,
isolated, Local Group galaxies, then we will know that the MW's tidal field is 
sufficient to drive global modes of star formation in its satellites.

There are three scenarios in which it might be acceptable to neglect 
the LMC-SMC interaction in this modeling.
First, our model requires eccentric orbits, a changing
tidal force, to produce bursts. Therefore, if the SMC-LMC orbit is fairly
circular, more circular than found by \cite{gardiner96}, we would not 
expect tidal triggering. Second, if the orbit has a period
that is smaller than the bin size (the resolution of our SAF), then we will not resolve the
bursts caused by the interaction and the LMC-SMC TT mode will simply appear to 
cause an elevated constant star formation rate. Third, if the bursts are associated only
with LMC/SMC pericenter passages that correspond to MW/SMC pericenter passages,
as appears to have happened a few hundred million years ago, then our orbital 
model would have the correct behavior but our interpretation would be incorrect.  Even 
though we cannot currently rule out the
LMC as the dominant factor in triggering bursts, 
the correspondence of both SMC bursts to MW pericenter
passages is strong circumstantial evidence for the importance of the MW passages on 
the SFR. Even if the LMC is eventually found to be the dominant source of triggering,
an analysis examining the MW's influence could place upper limits on the importance
of interactions.
Resolving this ambiguity is critical to further progress in this type of analysis.

\begin{figure}
\plotone{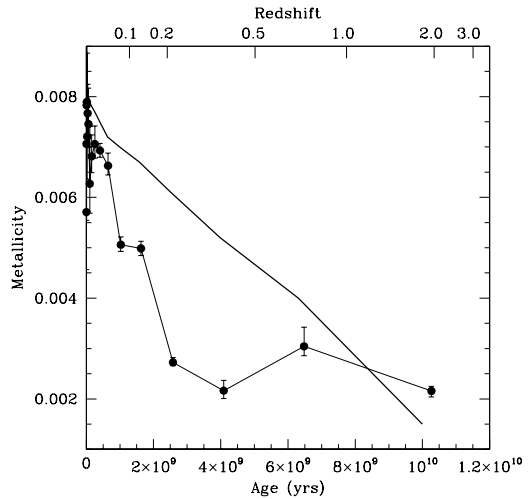}
\caption{Comparison of the chemical enrichment history 
has derived from a tidally-triggered model of star formation
(thick solid line; Model 2 in Table 1), and the derived
chemical enrichment history from \cite{harris03} (thin
solid line and data points). Once again the temporal position of the 
oldest point is somewhat ill-defined (see text).
\label{chem1}}
\end{figure}

\subsection{The Failings of Closed-Box Evolution and the Necessity for Infall}

Certain features in the SAF, namely the two peaks at $t < 4$ Gyr
are reproduced quite well with simple closed-box models, but other
features are not reproduced.
The closed-box models have two fundamental deficiencies: (1) they do not reproduce
the steep drop in the number of stars formed in the oldest and
second oldest bin (alternatively described as the relatively quite period
between 8 and 4 Gyr) and (2) they do not reproduce the flat, 
or perhaps even declining chemical enrichment history between 8
and 4 Gyr (Figure \ref{chem1}). Both of these shortcomings
suggest that these models are
not entirely satisfactory and that a possible solution may involve
infall of low-metallicity gas into a galaxy that had significantly
depleted its initial gas. The early exhaustion of gas would help explain the 
low SFR at $~$8 Gyr. The subsequent infall is necessary to dilute
what would otherwise be a fairly metal rich gas (in a closed-box
model the chemical abundance approaches the yield as the gas
is exhausted) and provide the fuel necessary to explain the star
formation at younger ages. Outflow might also be invoked, although
it is difficult to understand why outflow might mitigate chemical
enrichment at earlier times but not at later times unless the total mass
of the galaxy has changed significantly in the last 8 Gyr.

\begin{figure}
\plotone{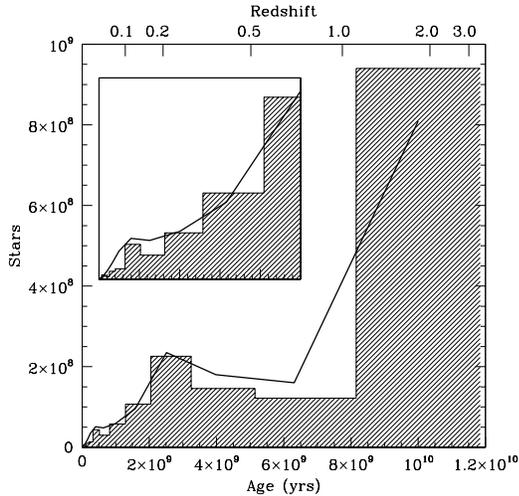}
\caption{Comparison of the tidally-triggered model
of star formation with infall that fits the SAF well
(thick solid line; Model 8 in Table 1) and the measured star formation
(histogram).
The inset is an expanded version of the most recent 2.5 Gyr.
\label{bestinfall}}
\end{figure}

In the spirit of continuing only with fairly simple models, we 
model infall as a single event of gas mass $M_g$ at time $t_{infall}$.
We now automate the exploration of parameter space by
defining a $\chi^2$ goodness-of-fit parameter, using the internal
statistical uncertainties in the SAF, and identifying the minimum $\chi^2$ value
and likelihood contours using $\Delta \chi^2$. Table
2 presents the range of parameter space probed, the best fit values, 
and the 90\% confidence intervals. The parameter space was
sampled coarsely resulting in some cases where the 90\%
interval includes only the best fit value.
The fitted values in this Table should be viewed with some
skepticism because none of the models is formally a good fit (the
best-fit $\chi^2_{\nu} = 3.1$, which is significantly better than any model in 
Table 1, but still not $\sim 1$).
The failure to statistically fit the data is in part due to
the exclusion of systematic errors (see \cite{harris03}) and the
oversimplification of the models. As an example of what we conclude
is an underestimation of the SAF uncertainties we show in Figure \ref{bestinfall}
a model (model 8 in Table 1) that fits the SAF quite well, but is not within the acceptable
ranges in Table 2.

\begin{figure}
\plotone{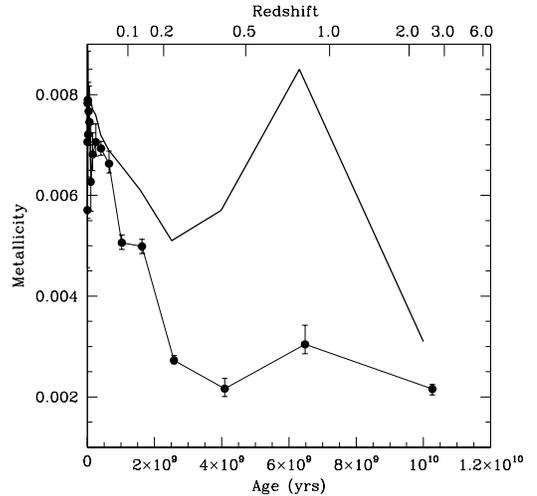}
\caption{Comparison of the chemical enrichment history
of the  tidally-triggered model
of star formation with infall that fits the SAF well
(thick solid line; Model 8 in Table 1) and the derived
chemical enrichment history from \cite{harris03} (thin
solid line and data points). 
\label{bestchem}}
\end{figure}

There seems to be little improvement
possible in reproducing the SAF, in particular because the lull in star formation at
intermediate ages is now fit quite well. Unfortunately, the chemical
enrichment history is not a good fit (Figure \ref{bestchem}, corresponds to model 8). 
Although this model reproduces the see-saw pattern at old ages, the
chemical abundance at those times is significantly higher in the models than what is 
observed. This discrepancy has three possible interpretations 1) the
model is missing a key ingredient (possibly outflow of high metallicity
gas in the early SMC), 2) the model is sufficiently sensitive
to the parameter choice that another models within the confidence intervals for
reproducing the SAF produces
a much better fit to the age-metallicity relationship
or 3) the observed chemical abundances are wrong
(recall that although we believe them to be accurate 
because of the agreement with abundances published for
individual clusters they are measured from broad band colors).  
We will ignore the latter possibility, although spectroscopic
chemical abundances of a large sample of stars in the SMC would be of 
great benefit in testing our derived age-metallicity relationship for field SMC stars.

\begin{figure}
\plotone{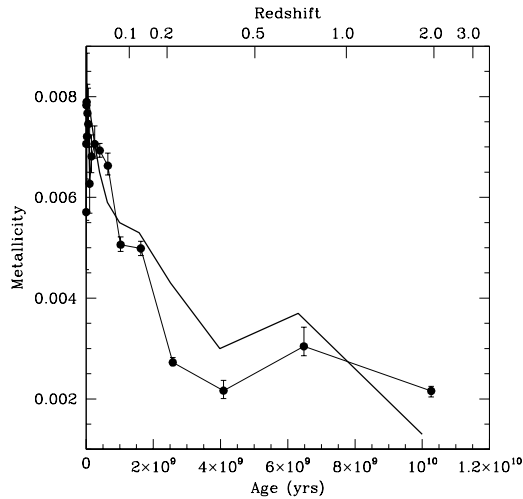}
\caption{Comparison of the chemical enrichment history 
derived from a tidally-triggered model of star formation with
infall
(thick solid line; Model 9 in Table 1), and the derived
chemical enrichment history from \cite{harris03} (thin
solid line and data points). Note that this particular model generates a
qualitatively good fit to the observed relationship.
\label{chem2}}
\end{figure}

\begin{figure}
\plotone{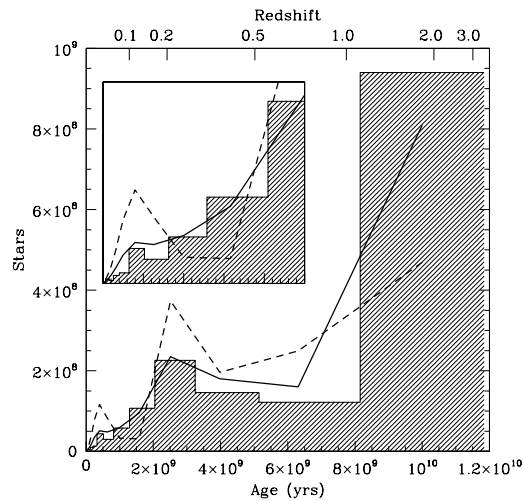}
\caption{Comparison of the star formation history 
derived from a tidally-triggered model of star formation
shown in Figure \ref{chem2}
(dashed line; Model 9 in Table 1), 
the tidally-triggered model with infall that fits
the star formation history better (solid line; Model 8 in Table 1)
and the SMC SAF (histogram).
\label{infall2}}
\end{figure}

We focus on the second of the three possibilities, again because we are
attempting to find the simplest model that is satisfactory.
We do find a model in which the
predicted abundance history is in better agreement with the data (Figure \ref{chem2})
at the expense of the agreement with the
SAF (Figure \ref{infall2}). 
Given that we can reproduce the age-metallicity
relation with one set of parameters and the star-formation history
with a slightly different set in this simple model, we conclude that
some combination of a slightly more complex model and the potential
systematic uncertainties in both the star formation history and 
chemical abundance history ensure that a fully satisfactory
solution exists within the framework advocated here. Given the uncertainties, the 
simplicity of the current model,  and the limited number of 
observational constraints, we do not think it worthwhile to push the data further
in addressing this issue or determining parameter values.
Nevertheless, there are several conclusions that are not sensitive to 
whether one favors the model that is the best fit to the SAF or the one that
is a good fit to the chemical
enrichment history. All models that fit the SAF require concentrated bursts
of star formation
and a significant infall event ($\sim$ 0.5 gas mass fraction) at intermediate
(4 -- 5 Gyr) ages. The models with stronger bursts are better at reproducing the age-metallicity
relationship while those we weaker bursts are better at reproducing the SAF. 
Unfortunately, with only one galaxy for comparison the models are
underconstrained and therefore not necessarily unique.

\subsection{Other Local Group Systems}

Although there are tens of LG systems with which such an model could ultimately
be tested, the data for most of these are not quite at the standard of that 
discussed here for the SMC. Published star formation histories do exist for
most of the dwarf spheroidals and for selected fields in the LMC, but
these are not done in a homogeneous manner  with the latest analysis
tools. An exception to this statement is the study of \cite{dolphin02} that attempted
to analyze a set of galaxies in a systematic way.
Even so, that work includes the caveat 
that the SFHs are derived from HST observations, which do not include
the entire galaxy. The presence of population gradients  in even these 
simple systems \citep{harbeck} makes necessary
larger area surveys to ensure 
a complete sampling of the SFH.
Orbital solutions using precise proper motion measurements are still quite rare 
and available in the literature only for a few of the dwarf spheroidals. 
Therefore, although we will
eventually be able to test the model on a large number of systems there
are none currently that provide as good a set of constraints as the SMC.

\section{Summary}

From our highly simplified modeling of the star formation history
of the Small Magellanic Cloud we conclude the following:

\noindent
1) The coincidence of two peaks in the SMC's stellar age function with the 
times of closest approach between SMC and MW, as determined from
proper motion measurements and the integration of the orbit within the gravitationally
potential of the Milky Way, suggest that a tidally-triggered mode of star
formation is important in the SMC. The principal shortcoming of our modeling
is the inability to model the effect of the LMC on the SMC due to our lack of detailed orbital
constraints of the Magellanic Clouds about each other at times $>$2 Gyr.
A composite model, where the more recent burst is due to the LMC, while the older
is due to the MW, is consistent with the current data, but adds too much complexity to 
constrain further with the current data.

\noindent
2) The elevation in the star formation rate is significant ($>$ factor
of 2) over the quiescent rate of star formation in the SMC. While the
tidally triggered 
mode is measurable at recent times, it may contribute as few as  
$10$\% of all stars in the SMC or more than 50\%. The uncertainty
in this number is due to the low resolution in our star formation history, particularly
at ages $>$ 2.5 Gyr. 

\noindent
3) The preferred values of the exponent for the radial dependence of tidally-triggered
mode of star formation suggest that tidally-triggered star formation
can be accurately modeled as an impulsive effect on the star formation rate when
the time resolution of the models is greater than about a hundred Myr.

\noindent
4) The failure of the closed-box model to reproduce both the stellar age
function and chemical abundances at intermediate ages in the SMC (4 $-$ 8 Gyr)
was resolved to a large degree with a single massive infall event that involved
about 50\% of the gas mass of the SMC infalling about 4 Gyr ago. We did not
show that this is a unique solution to the problems posed by the closed-box
model, but it is consistent with the differing distributions of younger and older
stars in the SMC \citep{zar00}.

\noindent
5) Conclusions 2-4 depend only on the assumption that the enhancements
observed in the stellar age function are due to periodic, tidally triggered star formation, but
do not depend on whether those bursts are driven by the LMC, MW, or a combination of the
two. If refinements in the orbital solutions of either the MW or LMC are able to exclude
one or the other as driving influences of the observed bursts, then limits could be placed on the importance of tidal interactions on the star formation rate.

Future comparisons of such models should compare directly to the color-magnitude
diagrams. It is therefore critically important for progress in this field to have a homogeneous,
well-documented public database of photometry of Local Group galaxies with the ancillary
data necessary to synthesize observed color-magnitude diagrams.

\section{Acknowledgments}

DZ acknowledges financial support from National Science Foundation
CAREER grant AST-9733111 and a fellowship from the David and Lucile
Packard Foundation. 

\vfill\eject

\begin{deluxetable}{lrrrrrrrrr}
\tablewidth{0pt}
\tablecaption{Model Runs}
\tiny
\tablehead{\colhead{Run No.}&\colhead{$b$}&\colhead{$t_{peri}$}&\colhead{P}&
\colhead{Age}&\colhead{$A$}&\colhead{$t_{infall}$}& \colhead{$f_{infall}$}
&\colhead{TT fraction}&\colhead{$\chi^2_\nu$}\\}
\startdata
1&...&...&...&17.0&0.0&...&...&...&11.6\\
2&2.4&0.2&2.4&17.0&100.0&...&...&0.93&9.4\\
3&2.4&0.2&1.5&17.0&100.0&...&...&0.93&12.1\\
4&2.4&2.1&2.4&17.0&100.0&..&...&0.93&11.9\\
5&10.0&0.2&2.4&17.0&100.0&...&...&0.38&116.5\\
6&2.4&0.2&2.4&17.0&10.0&...&...&0.58&6.9\\
7&2.4&0.2&2.4&17.0&1.0&...&...&0.12&11.7\\
8&6.0&0.3&2.4&17.0&4.0&4.0&0.5&0.11&6.2\\
9&4.2&0.3&2.4&17.0&60.0&5.0&0.4&0.78&80.6\\
\enddata
\end{deluxetable}

\begin{deluxetable}{lrr}
\tablewidth{0pt}
\tablecaption{Infall Model Best-Fit Parameters}
\tiny
\tablehead{
\colhead{Symbol}& \colhead{Range}& \colhead{Acceptable Range}\\}
\startdata
A&0.2$-$10.0&1.2[1.0,1.6]\\
$b$&2.0$-$7.0&4.6[2.8,7.0)\\
$t_{AGE}$ (Gyr)&14$-$17&17[17,17)\\
$t_{infall}$ (Gyr)&3.0$-$7.0&4.0[4.0,4.0]\\
$f_{infall}$          &0.3$-$0.6&0.4[0.5,0.5]\\
\enddata
\end{deluxetable}

\end{document}